\documentclass[12pt]{iopart}
\usepackage{iopams}  
\usepackage{graphicx}

\begin{document}

\title[]{Feeding of the elliptic flow by hard partons}

\author{Boris Tom\'a\v{s}ik$^{1,2}$ and P\'eter L\'evai$^3$}

\address{$^1$Univerzita Mateja Bela, 97401 Bansk\'a Bystrica, Slovakia}
\address{$^2$Czech Technical University in Prague, FNSPE, 11519 Prague, Czech Republic
}
\address{$^2$MTA KFKI RMKI, Budapest, Hungary}
\eads{\mailto{boris.tomasik@umb.sk}}

\begin{abstract}
We propose that in nuclear collisions at the LHC the elliptic flow may get a contribution
from leading hard and semihard partons which deposit energy and momentum into 
the hydrodynamic bulk medium. The crucial effect is that these partons induce wakes 
which interact and merge if they come together. The contribution to the integrated elliptic flow
is estimated with the help of a toy model to about 25\% of the observed value and 
shows strong event-by-event fluctuations.
\end{abstract}



\section{Introduction}
\label{intro}

The azimuthal anisotropy of hadron production in ultrarelativistic nuclear collisions, 
known under the term \emph{elliptic flow}, has proved to be an important observable
\cite{Ollitrault:1992bk,Voloshin:1994mz,Kolb:2000sd}. 
In particular, its observed size lead to conclusions that concern the time of thermalisation 
\cite{Heinz:2001xi,Bozek:2010aj}
and the influence of shear and bulk viscosity
\cite{Hirano:2005xf,Drescher:2007cd,Romatschke:2007mq,Teaney:2009qa,%
Masui:2009pw,Song:2009rh}. 
To get to these important and interesting 
characteristics of the bulk matter one must, however, reasonably quantify the 
influence of other effects
\cite{Luzum:2008cw,Bozek:2009dw,Shen:2010uy}. 
Among them there is our ignorance about the actual initial 
conditions for the hydrodynamic evolution of the fireball as well as the role of 
fluctuations in the initial conditions \cite{Andrade:2006yh,Schenke:2010rr,Petersen:2010md}. 
Also, the transition from dense hydrodynamic
matter to more dilute hadronic gas, which could slip out of equilibrium and later 
freezes-out, may also impact the momentum anisotropy
\cite{Petersen:1900zz,Hirano:2010jg,Hirano:2010je,Song:2011hk,Song:2011qa}. 
In this paper we describe a contribution to the elliptic flow which may be important at the LHC energy and possibly 
also at RHIC. 

The standard interpretation  of the elliptic flow is that it is caused by the anisotropy of the 
pressure gradients within the excited matter. They are larger in the direction of the impact 
parameter, which is usually called `in-plane' direction. (The other transverse direction 
is denoted `out-of-plane'.) The anisotropic pressure gradients lead to different accelerations 
of the collective flow in the different directions. The fireball then finally expands faster in 
the in-plane direction. Therefore, more hadrons are emitted in this direction and their transverse
momentum spectra are flatter  than in the out-of-plane direction. The measured size of this  
anisotropy, its dependence on the transverse 
momentum and on particle species depend on many effects. The one 
among them we want to focus at 
is the dependence on viscosity, since one would like to extract 
this transport coefficient from the data \cite{Shen:2010uy}.

Thus, if  there is an additional cause for the azimuthal anisotropy of hadron production, this might 
influence the statements
concerning fast thermalisation and maybe even modify the conclusion 
about low viscosity. In this paper, we consider such a mechanism. 

At the LHC, jets and minijets are produced copiously. 
Their energy loss when crossing the deconfined matter is so huge that only a few of them appear 
as distinguished jets
\cite{Adams:2003im,Aad:2010bu,Chatrchyan:2011sx,Aamodt:2010jd}. 
They rather transfer their energy and momentum into the bulk matter and initiate
collective phenomena there
\cite{Satarov:2005mv,CasalderreySolana:2004qm,Ruppert:2005uz,Renk:2005si,%
Neufeld:2008hs,Neufeld:2010tz,Betz:2008ka,Shuryak:2011vf,%
HwaPLB,HwaPRC}.  
Momentum must be conserved and thus the momentum of the hard parton 
must be transformed into momentum of a stream of matter (a wake), a Mach cone wave, or something like this. Much 
interest is currently devoted to such effects. 

\begin{figure}
\begin{center}
\includegraphics[width=0.6\textwidth]{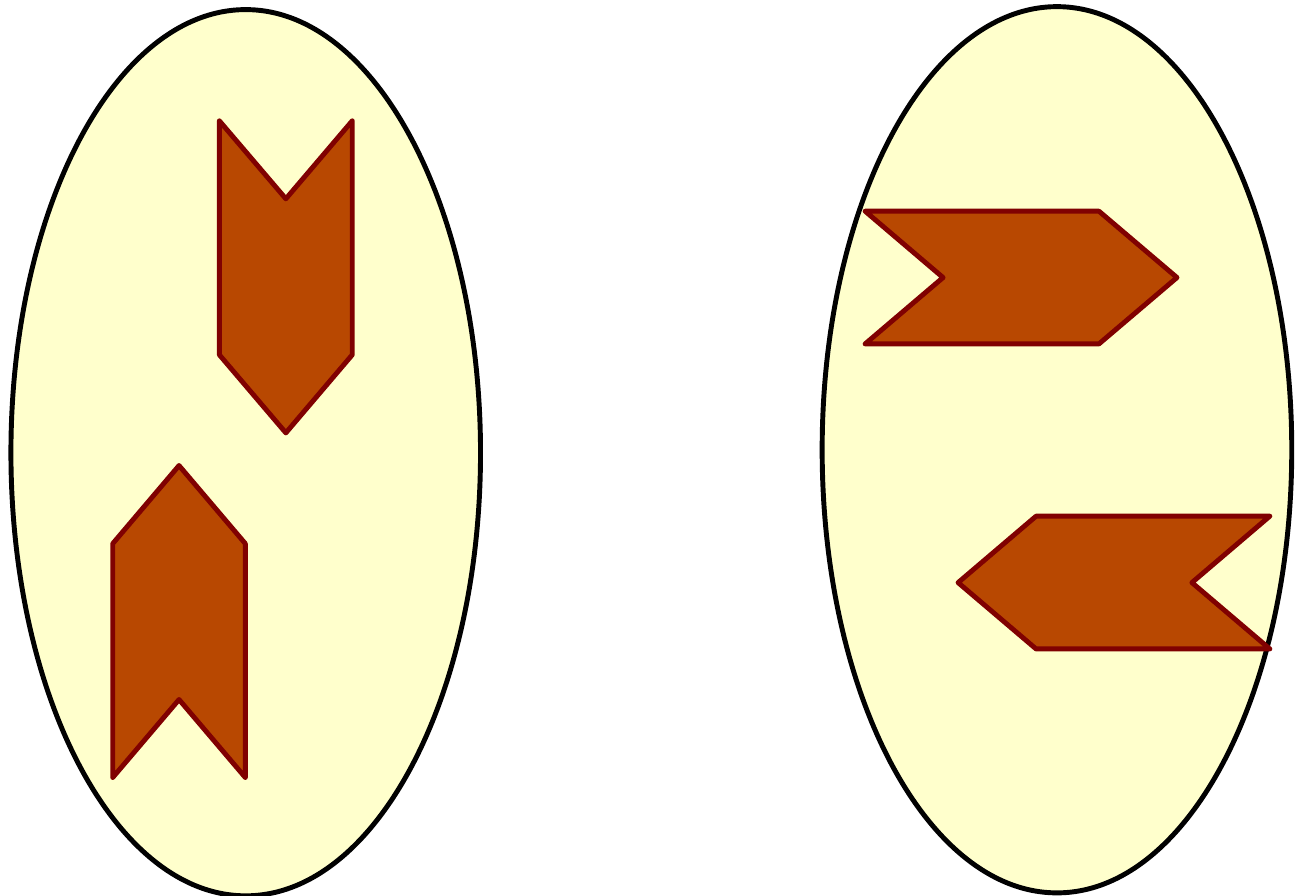}
\end{center}
\caption{%
Illustration of the likeness or unlikeness of two streams to meet. Left: two streams 
flowing in the out-of-plane direction are likely to meet. Right: as the fireball 
is elongated out of the reaction plane, two streams which flow in the in-plane direction 
have more space to pass each other without merging.
\label{f:ill}}
\end{figure}

A question arises, what would be the result of many such \emph{streams} if they all are initiated in the 
fireball? Some of these streams could merge and either 
cancel or flow in a new direction so that energy and momentum 
are conserved. Original hard partons are produced with no preferred transverse direction. The 
first expectation would be that the large number of fluid streams they initiate cancel out in 
some way and in the end there remains just thermalised matter with some energy density and no 
macroscopic flow.  However, in non-central collisions the argument might not be so straightforward. 
The streams have random directions, but their spatial distribution 
is \emph{not} isotropic, since it is given by the initial collision geometry. 
In the in-plane direction the fireball is narrower. Thus there is a good chance
that two streams having finite width and flowing in the out-of-plane direction will meet
(see Figure~\ref{f:ill}). 
On the other hand, 
streams could more easily pass each other without interacting when produced in the in-plane direction.
Thus---very grossly---streams perpendicular to the reaction plane cancel each 
other while those flowing in 
directions parallel to the reaction plane survive. As a result, the flow of bulk matter receives 
some feeding from hard partons and this feeding exhibits signs of a positive elliptic flow. 

Note that the scenario proposed here is related in its spirit to hydrodynamic 
simulations with hot spots in the initial conditions which are simulated on the 
event-by-event basis \cite{Andrade:2006yh,Schenke:2010rr,Holopainen:2010gz,%
Petersen:2010zt,Borysova:2011bn,Gardim:2011qn}.
Contribution to elliptic flow of the bulk from semi-hard partons has also been considered 
in \cite{HwaPLB,HwaPRC}.

Obviously, this argument is very schematic and to make it more sound one would need 
to integrate over all possible directions of the streams.  
The proper way to do so would be the use 
of a hydrodynamic simulation with included energy and momentum deposition from hard partons. 
At present, such simulations are not available due to technical complexity. Therefore, in order 
to obtain simple estimates of the possible size of the effect, we rather construct a 
simple toy model, which resembles the effect. 

In our toy model which is introduced in the next 
Section, we represent the flowing streams by flying blobs of matter with specified size. 
If two blobs meet, they merge into a heavier blob which moves so that 
energy and momentum are conserved. Originally, many blobs are generated. Their number 
and momenta correspond to the expected number and momenta of hard partons.  In the end, 
after merging of all blobs which should have merged, they all evaporate pions. In 
Section~\ref{elflow} we then analyse the distributions of these pions. It turns out that 
about 25\%  of the observed integrated elliptic flow may be due to our effect. We also find 
that this contribution is very strongly fluctuating. Conclusions are summarised in 
Section~\ref{conc}.


\section{The toy model}
\label{toymodel}

Streams within the fluid are represented by blobs. For each blob first its four-velocity is generated.
The momentum is generated according to the calculated 
distributions of the produced hard partons in transverse momentum and rapidity. For rapidity distributions 
at LHC and RHIC we assume that they are uniform in the central two units of rapidity. 

\subsection{Momentum distribution}

Transverse momentum spectra have been calculated and
the differential cross section for gluon production in proton-proton collision was 
parametrised as
\begin{equation}
\label{ptsig}
E\frac{d\sigma_{NN}}{dp^3} = \frac{1}{2\pi} \frac{1}{p_t} \frac{d\sigma_{NN}}{dp_t\, dy} = 
\frac{B}{\left ( 1 + p_t/p_0\right )^n}\, ,
\end{equation}
where $p_0$, $B$, and  $n$ are parameters. The parametrisation works fine 
in the $p_t$ interval from 2.5 to 12~GeV/$c$. Calculated spectra
deviate from this parametrisation for higher $p_t$. Note, however, that production of 
jets at such high $p_t$ is rare and thus does not contribute much to the total yield
and can be assumed to have small effect on the bulk when large number
of collisions is analysed. For a simulation at LHC energies we chose 
$B = 14.7$~mbarn/GeV$^2$, $p_0 = 6$~GeV, and $n= 9.5$.

Now we calculate the total number of blobs in a non-central symmetric collision of two nuclei with mass numbers
$A$ at the impact parameter $b$ ($b = |\vec b|$).
The cross-section for the production of the leading particle with $p_t$ larger that $p_m$ 
is then obtained by integrating eq.~(\ref{ptsig}) 
\begin{equation}
\sigma(p_m) = \int_{p_m}^{\infty} \int_{y_{\rm min}}^{y_{\rm max}} 
\frac{d\sigma_{NN}}{dp_t\, dy} \, dp_t\, dy\,  .
\end{equation}
The mean total number of leading particles with $p_t>p_m$ is then
\begin{equation}
N_j(p_m,b) = \frac{A^2 \, T_{AA}(b)\, \sigma(p_m)}{1 - \left (  1 - T_{AA}(b) \sigma(p_m) \right )^{A^2}}\, .
\end{equation}
In the last equation we introduced the overlap function
\begin{eqnarray}
T_{AA}(b) & = & \int_{\rm overlap}  T_A(\vec r)\, T_A(\vec r - \vec b)\, d^2\vec r \, ,
\\
T_A(\vec r) & = & 2\, \rho_0\, \sqrt{R_A^2 - r^2} \, .
\end{eqnarray}
where $T_A(\vec x)$ is the nuclear thickness function. The radius of the nucleus 
is $R_A$  and for the sake of our estimates we have assumed very simple profile
with constant nuclear density $\rho_0$.

\subsection{Evolution of blobs}

The blobs represent streams of bulk matter and  carry the momentum of the leading partons.
Once the momentum of the blob is given, for the simulation we need to determine the velocity
of the blob. It will be close to $c$ since all energy of the partons is basically
due to momentum. Technically, we choose a very small off-shell mass of $m=1$~GeV. Then, the velocity of a blob with transverse momentum $p_t$ and rapidity $y$ is
\begin{equation}
v^\mu =( m_t\, \cosh y,\, p_t\,  \cos\phi,\, 
p_t\, \sin\phi,\, m_t\,  \sinh y)\, ,
\end{equation}
where $m_t = \sqrt{p_t^2 + m^2}$ and the azimuthal angle $\phi$ is generated randomly from a uniform distribution. 

Jets are produced in pairs which are roughly back to back in the transverse plane
(but not longitudinally) so that the total transverse momentum vanishes. 
However, in the parametrisations of the leading parton differential cross section we have assumed 
some broadening of the transverse momentum due to intrinsic 
$\langle k_t^2 \rangle$. Since this is an initial state effect, it follows that the total 
transverse momentum of the hard parton 
pair should be of the order $\langle k_t^2\rangle$. Following 
this geometry we find that the away-side parton is not directed precisely in the oposite way but 
may deviate from this direction by some angle $\alpha$ of the order
\begin{equation}
\alpha^2 \simeq  \frac{\langle k_t^2 \rangle}{p_t^2}\, . 
\end{equation}
Here, we have assumed that $\langle k_t^2 \rangle \ll p_t^2$. Thus the azimuthal angle of the generated away side leading parton 
shall deviate from the opposite direction by this $\alpha$, which will be generated from a Gaussian
distribution with the width $\sqrt{\langle k_t^2 \rangle/p_t^2}$. In case of small $p_t$ this may become a 
large number. Thus if $\langle k_t^2 \rangle/p_t^2>0.7$ we shall fix the width to 0.7.

The transverse positions at which the blobs start moving are distributed according to 
the density of binary nucleon-nucleon collisions
\begin{equation}
\rho_b(\vec r) = T_A(\vec r)T_B(\vec r - \vec b)
\end{equation}

The time at which the blob is created is delayed by some formation time $\tau_0$ (set usually to 0.6~fm/$c$) 
multiplied by $\gamma$. The longitudinal position is then $\tau_0v_3$. 
Longitudinal rapidity is chosen from a uniform distribution. 

Two blobs merge when they approach each other in their pair CMS closer than $2R_b$, where 
$R_b$ is a model parameter which we will vary in our simulations. In this formulation it is the size of a blob
but it actually represents the radius of the fluid stream. In the merger a new blob with the same size 
and a higher energy content is created. It will move with velocity $v$ and the direction of the velocity 
is chosen such that energy and momentum are conserved. 

When there are no more mergers the blobs evaporate into pions. In the rest frame  of the blob, 
pions emitted by that blob are distributed thermally  with a kinetic temperature $T$.
In our simulation we choose T=160~MeV.


\section{Elliptic flow}
\label{elflow}

The elliptic flow coefficient $v_2$ is defined as the second order Fourier coefficient in 
the decomposition of the azimuthal single-particle distribution 
\cite{Voloshin:1994mz}
\begin{equation}
P(\phi) = \frac{dN}{d\phi} = N_0 \left ( 1 + 2v_2\, \cos[2(\phi - \phi_0)] + \dots \right ) \, , 
\end{equation}
where $\phi_0$ is the angle of the event plane. We 
have written this relation for midrapidity so that certain symmetry constraints apply. 
Since in our simulation we know the orientation of the reaction plane, we can always 
use the coordinate frame in which  $\phi_0 = 0$ and determine $v_2$ from 
\begin{equation}
v_2 = \frac{\int_0^{2\pi} P(\phi)\, \cos(2\phi)\, d\phi}{\int_0^{2\pi} P(\phi) \, d\phi}\, .
\end{equation}
We measure $v_2$ in each event as the average of $\cos(2\phi)$ over all particles and then 
we take the average over all events. 

\begin{figure}
\begin{center}
\includegraphics[width=0.49\textwidth]{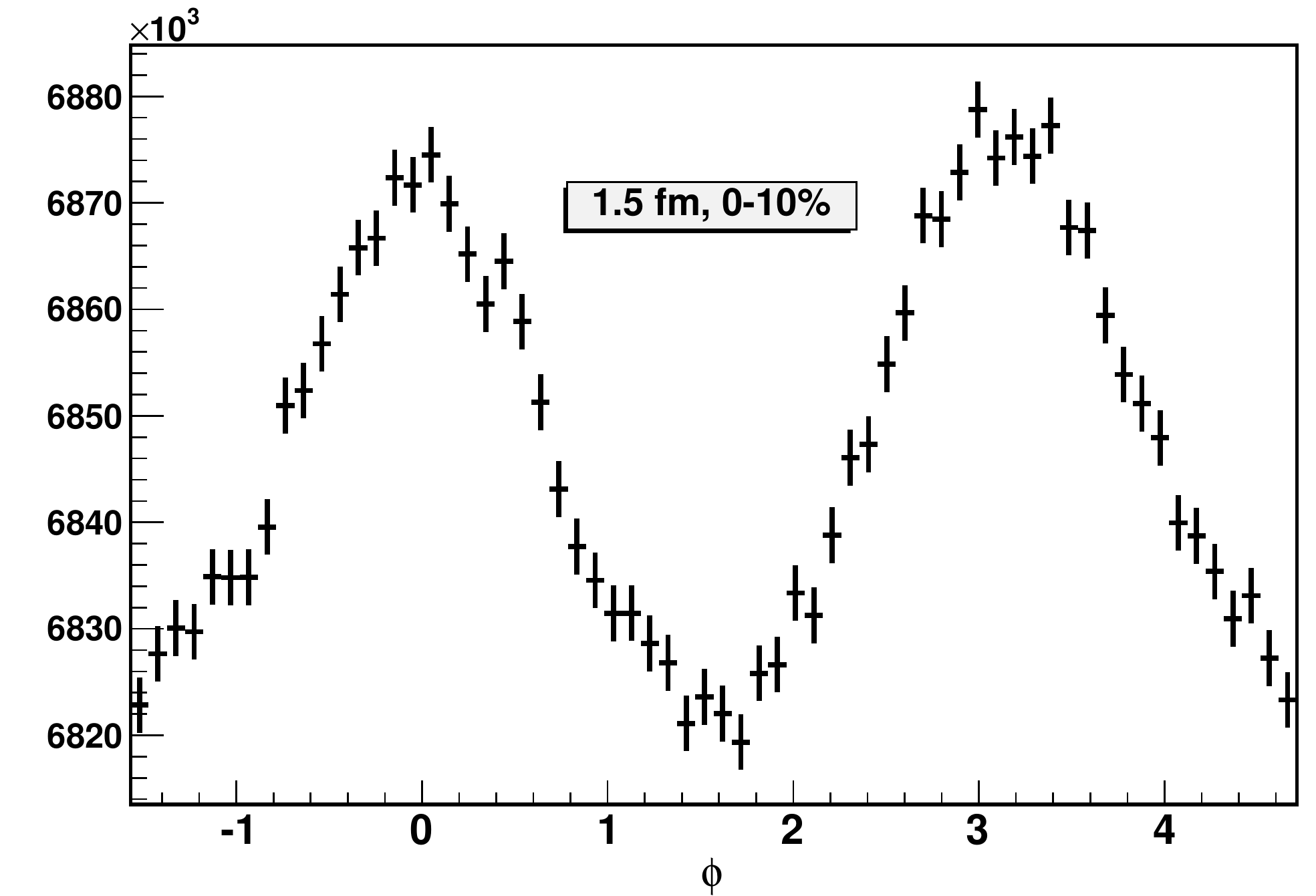}
\includegraphics[width=0.49\textwidth]{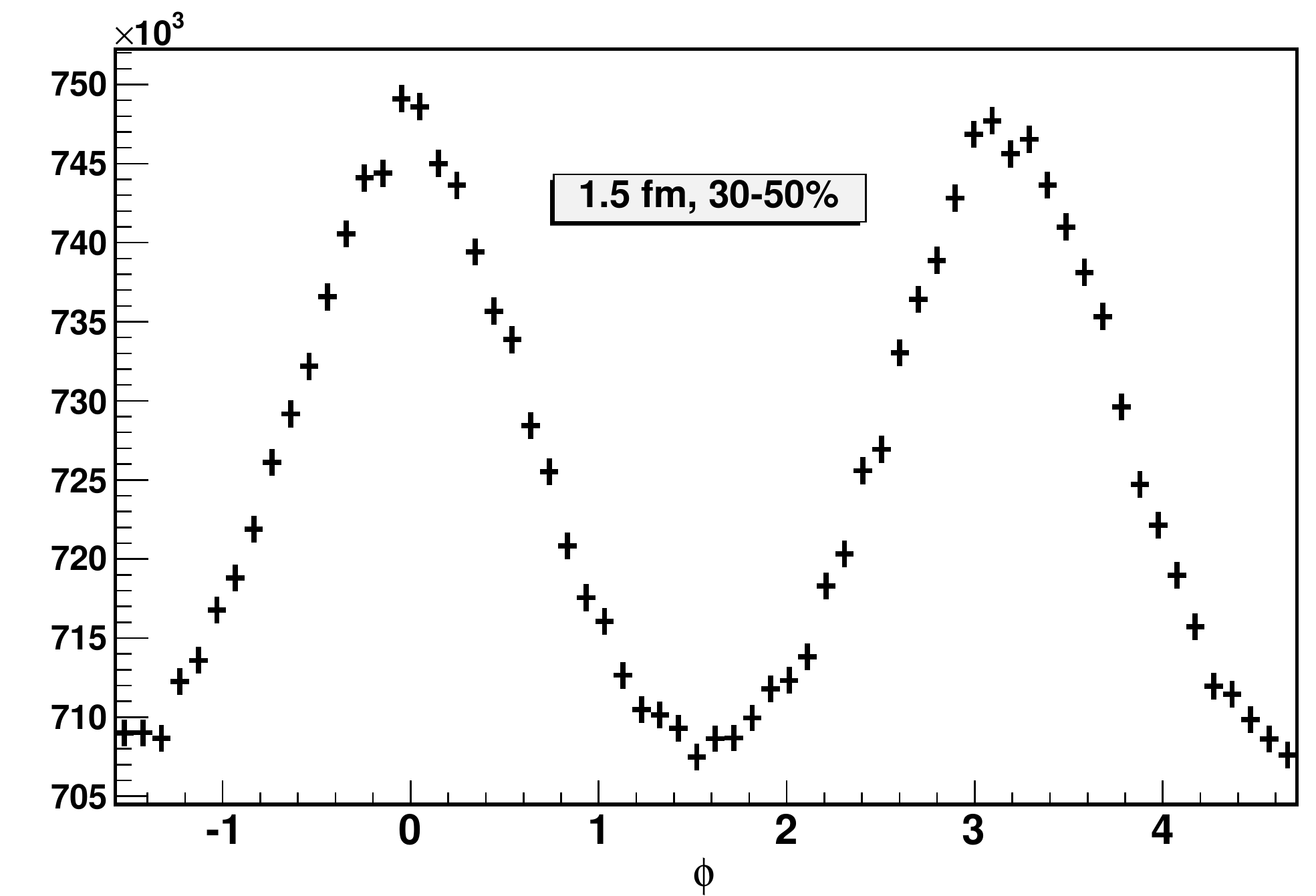}
\includegraphics[width=0.49\textwidth]{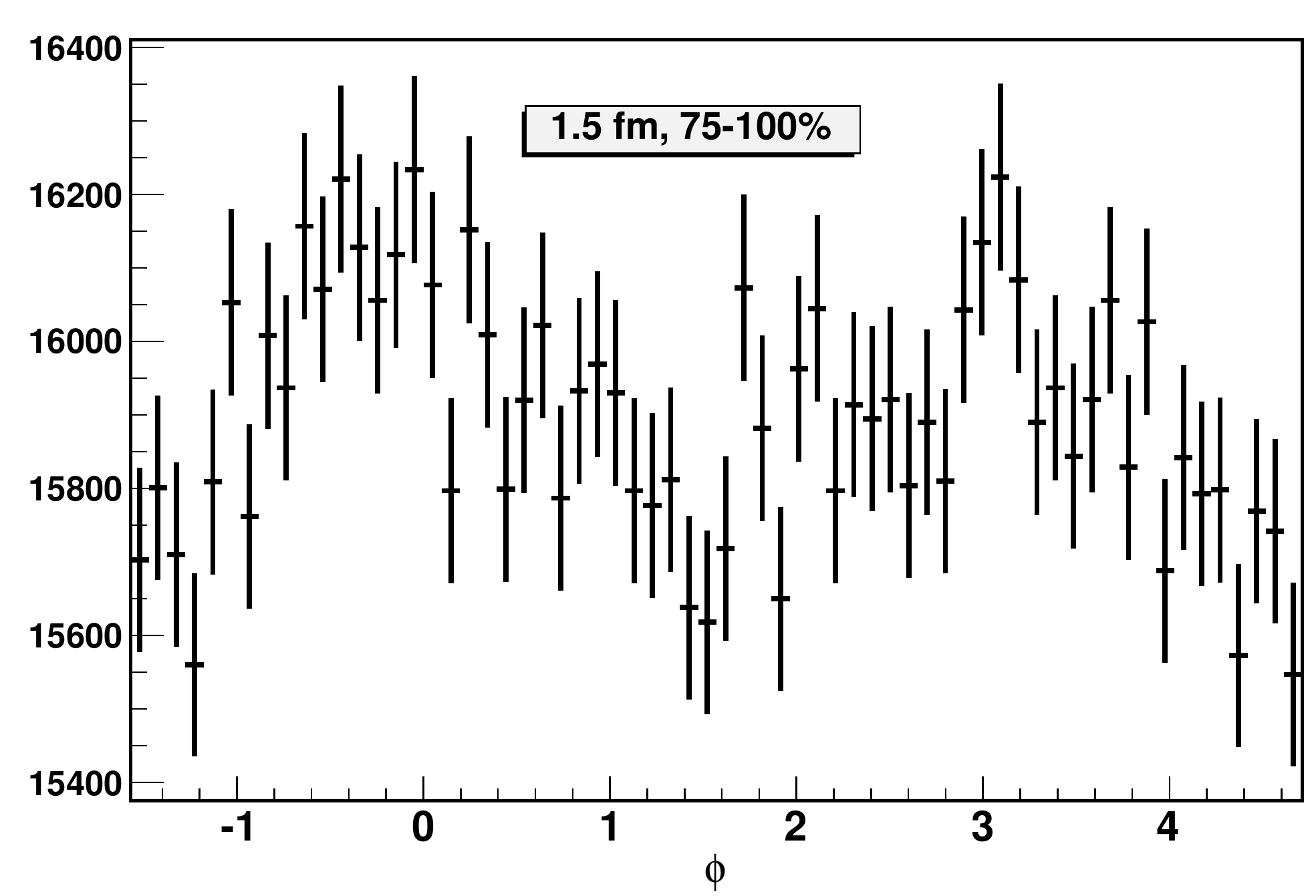}
\includegraphics[width=0.49\textwidth]{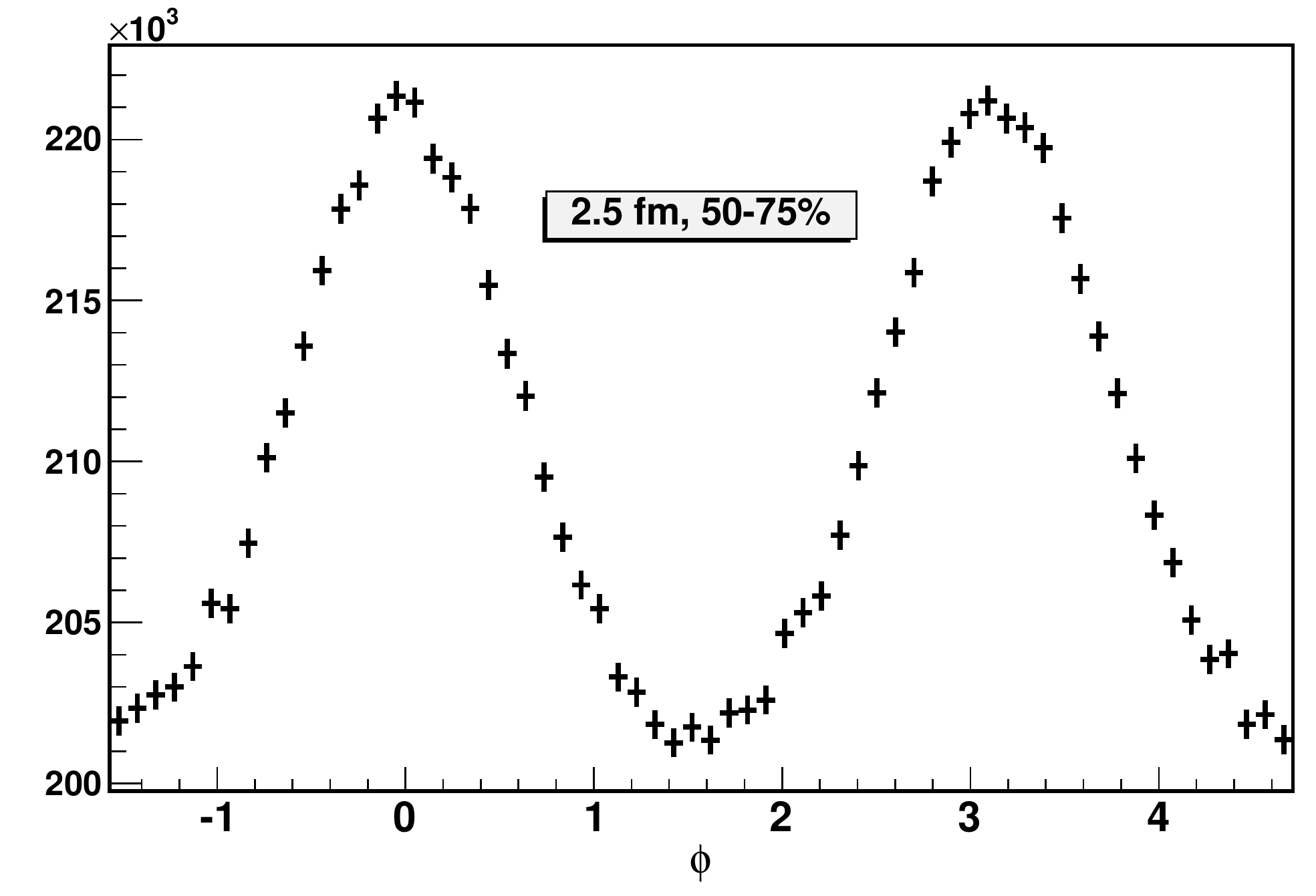}
\end{center}
\caption{   
Histograms of the azimuthal distributions of pions integrated over $p_t$. Different
panels correspond to different radii of the streams $R_b$
and different centrality classes, as indicated in the panels. 
\label{f:hist}}
\end{figure}

In Figure~\ref{f:hist}
we show the azimuthal distributions summed over all simulated events. For the most 
central classes we simulated more than 150,000 events and for centralities over 50\%
the samples count 100,000 events. For the analysis we accepted pions in rapidity 
window from --1 to 1. 

In the histograms we clearly see that the production of final state hadrons is correlated 
with the reaction plane. Thus we deal with true flow signal and not just with a non-flow 
effect stemming from correlation of individual hadrons with each other. The difference between 
results of simulations with two chosen radii of streams is not very large. 

\begin{figure}
\begin{center}
\includegraphics[width=0.7\textwidth]{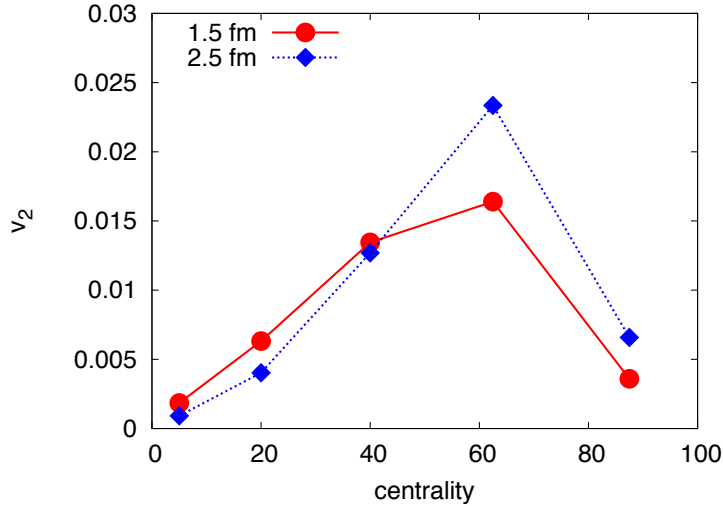}
\end{center}
\caption{   
Centrality dependence of $v_2$ for different radii of the streams: $R_b = 1.5$~fm
(solid red line), $R_b = 2.5$~fm (dotted blue line). 
\label{f:v2cent}}
\end{figure}

In order to better see how the anisotropy depends on centrality and the size of the streams
we plot in Figure \ref{f:v2cent}
the elliptic flow parameter $v_2$ as function of centrality. As expected, $v_2$
increases when departing from central events towards
more peripheral. This growth turns into 
decrease for the most peripheral collisions due to lower density of streams where 
the chance of mergers is smaller. The integrated $v_2$ in our simulations reaches 
up to about 0.02. Note that this amounts up to 25\% of the integrated $v_2$ measured 
in Pb+Pb collisions at 2.76 $A$TeV \cite{Aamodt:2010pa} 
and a similar share of the elliptic flow at RHIC \cite{:2008ed,Afanasiev:2009wq}. 
Thus we want to conclude that the proposed effect of momentum feeding from hard leading 
partons with subsequent stream interaction may contribute considerably to the observed
elliptic flow. 

We also found in the simulation that $v_2$ exhibits large fluctuations of the order of 
mean value or even more in some cases. This follows from the nature of the effect which 
crucially depends on the mergers of streams and vanishes if no such effect occur in a given event. The fluctuation might possibly be milder if complete hydrodynamic evolution is added 
to the picture. Nevertheless, we expect that one of the signatures of the effect discussed here 
will be enhanced $v_2$ fluctuations.


\section{Conclusions}
\label{conc}

Our study qualitatively shows that in nuclear collisions at the LHC relevant effect for 
momentum anisotropy may be the transfer of momentum from the leading hard partons 
to hydrodynamic medium. Crucial ingredient is the interaction of streams induced in the 
bulk medium. They may merge and flow together. The net effect is the deflection of some 
of the streams from the out-of-plane direction into the in-plane direction. 

The toy model simulation amounted to 25\% of the elliptic flow that is measured in 
Pb+Pb collisions  at the LHC. The conclusion is that the effect may be important and 
should be included in simulations aimed at extraction of bulk matter properties by 
comparing to experimental data. So far, simulations with one jet depositing momentum 
into bulk matter have been performed \cite{Betz:2008ka}. 
However, to our best knowledge no 
hydrodynamic simulations exist, where the effect of energy and momentum feeding 
from many hard partons would be investigated.

\paragraph{Acknowledgements}
In its initial phase, this work was supported by the Hungaro-Slovak collaboration grant 
No. SK-HU-029-06 (SK) and SK 20/2006  (HU). BT ackowledges support 
via grants Nos MSM~6840770039, and  LC~07048 (Czech Republic). This work was partially
supported by the Agency of the Slovak Ministry of Education for the Structural Funds of the EU, under project ITMS:26220120007. PL acknowledges the support of OTKA grant No 77816.



\section*{References}

\begin{thebibliography}{99}

\bibitem{Ollitrault:1992bk}
  J.~Y.~Ollitrault,
  Phys.\ Rev.\  D {\bf 46} (1992) 229.

\bibitem{Voloshin:1994mz}
  S.~Voloshin and Y.~Zhang,
  Z.\ Phys.\  C {\bf 70} (1996) 665
  [arXiv:hep-ph/9407282].

\bibitem{Kolb:2000sd}
  P.~F.~Kolb, J.~Sollfrank and U.~W.~Heinz,
  Phys.\ Rev.\  C {\bf 62} (2000) 054909
  [arXiv:hep-ph/0006129].

\bibitem{Heinz:2001xi}
  U.~W.~Heinz and P.~F.~Kolb,
  Nucl.\ Phys.\  A {\bf 702} (2002) 269
  [arXiv:hep-ph/0111075].

\bibitem{Bozek:2010aj}
  P.~Bo\.zek and I.~Wyskiel-Piekarska,
  Phys.\ Rev.\  C {\bf 83} (2011) 024910
  [arXiv:1009.0701 [nucl-th]].

\bibitem{Hirano:2005xf}
  T.~Hirano, U.~W.~Heinz, D.~Kharzeev, R.~Lacey and Y.~Nara,
  Phys.\ Lett.\  B {\bf 636} (2006) 299
  [arXiv:nucl-th/0511046].

\bibitem{Drescher:2007cd}
  H.~J.~Drescher, A.~Dumitru, C.~Gombeaud and J.~Y.~Ollitrault,
  Phys.\ Rev.\  C {\bf 76} (2007) 024905
  [arXiv:0704.3553 [nucl-th]].

\bibitem{Romatschke:2007mq}
  P.~Romatschke and U.~Romatschke,
  Phys.\ Rev.\ Lett.\  {\bf 99} (2007) 172301
  [arXiv:0706.1522 [nucl-th]].

\bibitem{Teaney:2009qa}
  D.~A.~Teaney,
  arXiv:0905.2433 [nucl-th].

\bibitem{Masui:2009pw}
  H.~Masui, J.~Y.~Ollitrault, R.~Snellings and A.~Tang,
  Nucl.\ Phys.\  A {\bf 830} (2009) 463C
  [arXiv:0908.0403 [nucl-ex]].

\bibitem{Song:2009rh}
  H.~Song and U.~W.~Heinz,
  Phys.\ Rev.\  C {\bf 81} (2010) 024905
  [arXiv:0909.1549 [nucl-th]].

\bibitem{Luzum:2008cw}
  M.~Luzum and P.~Romatschke,
  Phys.\ Rev.\  C {\bf 78} (2008) 034915
  [Erratum-ibid.\  C {\bf 79} (2009) 039903]
  [arXiv:0804.4015 [nucl-th]].

\bibitem{Bozek:2009dw}
  P.~Bo\.zek,
  Phys.\ Rev.\  C {\bf 81} (2010) 034909
  [arXiv:0911.2397 [nucl-th]].

\bibitem{Shen:2010uy}
  C.~Shen, U.~Heinz, P.~Huovinen and H.~Song,
  Phys.\ Rev.\  C {\bf 82} (2010) 054904
  [arXiv:1010.1856 [nucl-th]].

\bibitem{Andrade:2006yh}
  R.~Andrade, F.~Grassi, Y.~Hama, T.~Kodama and O.~J.~Socolowski,
  Phys.\ Rev.\ Lett.\  {\bf 97} (2006) 202302
  [arXiv:nucl-th/0608067].

\bibitem{Schenke:2010rr}
  B.~Schenke, S.~Jeon and C.~Gale,
  Phys.\ Rev.\ Lett.\  {\bf 106} (2011) 042301
  [arXiv:1009.3244 [hep-ph]].

\bibitem{Petersen:2010md}
  H.~Petersen and M.~Bleicher,
  Phys.\ Rev.\  C {\bf 81} (2010) 044906
  [arXiv:1002.1003 [nucl-th]].

\bibitem{Petersen:1900zz}
  H.~Petersen, J.~Steinheimer, G.~Burau and M.~Bleicher,
  Eur.\ Phys.\ J.\  C {\bf 62} (2009) 31.

\bibitem{Hirano:2010jg}
  T.~Hirano, P.~Huovinen and Y.~Nara,
  Phys.\ Rev.\  C {\bf 83} (2011) 021902
  [arXiv:1010.6222 [nucl-th]].

\bibitem{Hirano:2010je}
  T.~Hirano, P.~Huovinen and Y.~Nara,
  arXiv:1012.3955 [nucl-th].

\bibitem{Song:2011hk}
  H.~Song, S.~A.~Bass, U.~W.~Heinz, T.~Hirano and C.~Shen,
  arXiv:1101.4638 [nucl-th].

\bibitem{Song:2011qa}
  H.~Song, S.~A.~Bass and U.~W.~Heinz,
  arXiv:1103.2380 [nucl-th].

\bibitem{Adams:2003im}
  J.~Adams {\it et al.}  [STAR Collaboration],
  Phys.\ Rev.\ Lett.\  {\bf 91} (2003) 072304
  [arXiv:nucl-ex/0306024].

\bibitem{Aad:2010bu}
  G.~Aad {\it et al.}  [Atlas Collaboration],
  Phys.\ Rev.\ Lett.\  {\bf 105} (2010) 252303
  [arXiv:1011.6182 [hep-ex]].

\bibitem{Chatrchyan:2011sx}
  S.~Chatrchyan {\it et al.}  [CMS Collaboration],
  arXiv:1102.1957 [nucl-ex].

\bibitem{Aamodt:2010jd}
  K.~Aamodt {\it et al.}  [ALICE Collaboration],
  Phys.\ Lett.\  B {\bf 696} (2011) 30
  [arXiv:1012.1004 [nucl-ex]].

\bibitem{Satarov:2005mv}
  L.~M.~Satarov, H.~Stoecker and I.~N.~Mishustin,
  Phys.\ Lett.\  B {\bf 627} (2005) 64
  [arXiv:hep-ph/0505245].

\bibitem{CasalderreySolana:2004qm}
  J.~Casalderrey-Solana, E.~V.~Shuryak and D.~Teaney,
  J.\ Phys.\ Conf.\ Ser.\  {\bf 27} (2005) 22
  [Nucl.\ Phys.\  A {\bf 774} (2006) 577]
  [arXiv:hep-ph/0411315].

\bibitem{Ruppert:2005uz}
  J.~Ruppert and B.~M\"uller,
  Phys.\ Lett.\  B {\bf 618} (2005) 123
  [arXiv:hep-ph/0503158].

\bibitem{Renk:2005si}
  T.~Renk and J.~Ruppert,
  Phys.\ Rev.\  C {\bf 73} (2006) 011901
  [arXiv:hep-ph/0509036].

\bibitem{Neufeld:2008hs}
  R.~B.~Neufeld,
  Phys.\ Rev.\  D {\bf 78} (2008) 085015
  [arXiv:0805.0385 [hep-ph]].

\bibitem{Neufeld:2010tz}
  R.~B.~Neufeld and T.~Renk,
  Phys.\ Rev.\  C {\bf 82} (2010) 044903
  [arXiv:1001.5068 [nucl-th]].

\bibitem{Betz:2008ka}
  B.~Betz, J.~Noronha, G.~Torrieri, M.~Gyulassy, I.~Mishustin and D.~H.~Rischke,
  Phys.\ Rev.\  C {\bf 79} (2009) 034902
  [arXiv:0812.4401 [nucl-th]].

\bibitem{Shuryak:2011vf}
  E.~Shuryak,
  arXiv:1101.4839 [hep-ph].

\bibitem{HwaPLB}
R.C.~Hwa, Phys.~Lett.~B \textrm{666} (2008) 228.

\bibitem{HwaPRC}
C.~B.~Chiu, R.~C.~Hwa, C.~B.~Yang, Phys.~Rev.~C \textrm{78} (2008) 044903.

\bibitem{Holopainen:2010gz}
  H.~Holopainen, H.~Niemi and K.~J.~Eskola,
  Phys.\ Rev.\  C {\bf 83} (2011) 034901
  [arXiv:1007.0368 [hep-ph]].

\bibitem{Petersen:2010zt}
  H.~Petersen, C.~Coleman-Smith, S.~A.~Bass and R.~Wolpert,
  J.\ Phys.\ G {\bf 38} (2011) 045102
  [arXiv:1012.4629 [nucl-th]].

\bibitem{Borysova:2011bn}
  M.~S.~Borysova, I.~A.~Karpenko and Yu.~M.~Sinyukov,
  arXiv:1102.2084 [nucl-th].

\bibitem{Gardim:2011qn}
  F.~G.~Gardim, F.~Grassi, Y.~Hama, M.~Luzum and J.~Y.~Ollitrault,
  arXiv:1103.4605 [nucl-th].

\bibitem{Aamodt:2010pa}
  K.~Aamodt {\it et al.}  [The ALICE Collaboration],
 Phys.\ Rev.\ Lett.\ \textbf{105} (2010) 252302. 
 [arXiv:1011.3914 [nucl-ex]].

\bibitem{:2008ed}
  B.~I.~Abelev {\it et al.}  [STAR Collaboration],
  Phys.\ Rev.\  C {\bf 77} (2008) 054901
  [arXiv:0801.3466 [nucl-ex]].

\bibitem{Afanasiev:2009wq}
  S.~Afanasiev {\it et al.}  [PHENIX Collaboration],
  Phys.\ Rev.\  C {\bf 80} (2009) 024909
  [arXiv:0905.1070 [nucl-ex]].




\end{thebibliography}
\end{document}